\documentclass[review]{elsarticle}

\usepackage{lineno,hyperref}
\usepackage{bm}
\usepackage{amsmath}
\usepackage{amssymb}
\usepackage{graphicx}
\usepackage{upgreek}
\usepackage{multirow}
\usepackage[ruled,lined,commentsnumbered,boxed]{algorithm2e}
\usepackage{epstopdf}
\usepackage{epsfig}
\usepackage{psfrag}
\usepackage{subfigure}
\usepackage{array}
\usepackage{booktabs}
\usepackage{threeparttable}

\journal{Neural Networks}

\begin{document}

\begin{frontmatter}

\title{Hierarchical feature fusion framework for frequency recognition in SSVEP-based BCIs}


\author[mymainaddress,mythirdaddress,myfifthddress]{Yangsong Zhang}

\author[mysecondaddress,mysixthaddress]{ Erwei Yin\corref{mycorrespondingauthor}}
\cortext[mycorrespondingauthor]{Corresponding author}
\ead{yinerwei1985@gmail.com}

\author[mythirdaddress,myfifthddress]{Fali Li}

\author[myfourthaddress]{Yu Zhang}

\author[mythirdaddress,myfifthddress]{Daqing Guo}

\author[mythirdaddress,myfifthddress]{Dezhong Yao}

\author[mythirdaddress,myfifthddress]{Peng Xu\corref{mycorrespondingauthor}}
\ead{xupeng@uestc.edu.cn}

\address[mymainaddress]{School of Computer Science and Technology, Southwest University of Science and Technology, Mianyang  621010, China}
\address[mysecondaddress]{National Institute of Defense Technology Innovation, Academy of Military Sciences China, Beijing, 100081 China}
\address[mysixthaddress]{Tianjin Artificial Intelligence Innovation Center (TAIIC), Tianjin, 300450 China}
\address[mythirdaddress]{Clinical Hospital of Chengdu Brain Science Institute, MOE Key Lab for Neuroinformation, University of Electronic Science and Technology of China, Chengdu 610054, China}
\address[myfourthaddress]{Department of Psychiatry and Behavior Sciences, Stanford University, Stanford, CA 94305, USA}
\address[myfifthddress]{School of life Science and technology, Center for information in medicine, University of Electronic Science and Technology of China, Chengdu, 611731, China}

\begin{abstract}
Effective frequency recognition algorithms are critical in steady-state visual evoked potential (SSVEP) based brain-computer interfaces (BCIs). In this study, we present a hierarchical feature fusion framework which can be used to design high-performance frequency recognition methods. The proposed framework includes two primary technique for fusing features: spatial dimension fusion (SD) and frequency dimension fusion (FD). Both SD and FD fusions are obtained using a weighted strategy with a nonlinear function. To assess our novel methods, we used the correlated component analysis (CORRCA) method to investigate the efficiency and effectiveness of the proposed framework. Experimental results were obtained from a benchmark dataset of thirty-five subjects and indicate that the extended CORRCA method used within the framework significantly outperforms the original CORCCA method. Accordingly, the proposed framework holds promise to enhance the performance of frequency recognition methods in SSVEP-based BCIs.

\end{abstract}


\end{frontmatter}


\section{Introduction}
A brain-computer interface (BCI) is a type of communication system that can directly translate brain signals into digital commands for the control of external devices without the involvement of the peripheral nerves and muscles. BCI systems show great promise for providing communication access to both people with severe motor disabilities and typically developed individuals alike ~\cite{wolpaw2002,chaudhary2016brain,MBCILiyq2016}. Due to its relative portability and low cost, electroencephalography (EEG) remains the most widely investigated sensing modality in BCI research due in part to its excellent temporal resolution~\cite{chen2015PNAS}. To date, several types of EEG signals have been used to design and operate BCIs, e.g. ERP~\cite{JingJ2011,townsend2016,xu2018}, sensorimotor rhythm (SMR)~\cite{feng2018,Jiao2018,liuyy2018}, steady-state visual evoked potential (SSVEP)~\cite{MFSC2012,Maye2017,Oikonomou2018}, and motion-onset visual evoked potentials~\cite{guo2008,ma2017}, etc. In addition, some researchers have begun to explore hybrid BCI approaches using multiple signals and modalities~\cite{liyq2010,yin2015hybrid,xu2013,Edelman2018}. In recent years, an increasing number of researchers have focused on SSVEP-based BCIs because of their high information transfer rate (ITR) and minimal user training requirements ~\cite{Yin2015,chen2014,Jiao2017,zhang2018a,jiang2018}.

For SSVEP-based BCIs, an effective frequency detection algorithm plays an important role in overall system performance ~\cite{chen2015PNAS}. In literature, various techniques for SSVEP feature extraction and classification have been developed~\cite{friman2007,Zerafa2018}. Among them, the most popular are based on multivariate statistical algorithms, such as canonical correlation analysis (CCA)~\cite{Lin2007,chen2015filter}, and multivariate synchronization index (MSI)~\cite{MSI2014,ZhangRMSI2016}, etc, which are easy to implement without complex optimization procedures. Recently, we proposed a novel frequency recognition method, termed CORRCA~\cite{zhangys2018b}, based on the correlated component analysis (COCA) which is also a multivariate statistical algorithm~\cite{Dmochowski2012,dmochowski2014audience,CohenENEURO2016}. The CORRCA method significantly outperforms the state-of-art CCA method~\cite{zhangys2018b}. As we know, CCA is a multivariate statistical method that measures the correlation between two sets of signals~\cite{Hotelling1936}. It requires that the canonical projection vectors be orthogonal and it generates two different projection vectors for the two signals. In contrast, COCA aims to maximize the Pearson Product Moment Correlation coefficient and does not necessitate orthogonality between the projection vectors. Furthermore, it generates only one projection vectors for the two signals this simplifying subsequent analysis~\cite{Pearson1896}. For EEG signal analysis, the COCA method maybe more efficient and practical in real-world applications~\cite{zhangys2018b,Dmochowski2012}.

Although standard CCA and CORRCA algorithms have been applied in literature with satisfactory results, the performance of these system could be further improved by applying sophisticated signal processing technologies. One such approach is the filter bank analysis method~\cite{chen2015filter}, which employs several bandpass filters to generate multiple subband components of the input signals. This filter bank based approach first carries out a frequency recognition method to obtain features on each subband component, then integrates all the features in classification using a weighted combination. In fact, we can consider the filter bank technique as a form feature fusion in frequency domain. This fusion strategy is just one way to explore the discriminating information implicit in the original signals. Here we term it a frequency domain fusion (FD fusion). In a recent motor imagery based BCI study, it was determined that weighting and regularizing common spatial patterns (CSP) features allow for the use of preclude the need for feature selection thus avoiding the loss of valuable information, and enhancing the performance of CSP~\cite{mishuhina2018}. We can consider this type of feature weighted method as a feature fusion strategy in the spatial domain. Analogously, we term it the spatial domain fusion(SD fusion). Both the standard CCA method and CORRCA method provide multiple correlation coefficients to measure the correlation between two multi-dimensional signals. It is worth noting that traditionally only the largest coefficients is selected as feature while all others are ignored~\cite{Lin2007,chen2015filter,nakanishi2015,zhangys2018b}. This inevitably results in the loss of the discriminative information. We posit that merging these two fusion strategies will result in a more robust target detection framework for SSVEP-based BCIs. To the best of our knowledge, a framework including both SD and FD fusions has never been introduced in SSVEP-based BCI to date.

Motivated by previous studies, here we propose a hierarchical feature fusion framework for SSVEP target detection by implementing both SD and FD fusions. In order to evaluate the efficiency and effectiveness of the proposed framework, we utilized the CORRCA method as a representative frequency recognition algorithm to investigate if the proposed framework can affect its performance. Furthermore, we conducted extensive experimental evaluation on a benchmark dataset of thirty-five subjects.

The remainder of this paper is organized as follows. Section II describes the proposed framework as well as its implementation and application techniques; Section III describes the dataset and performance evaluation; Section IV presents the experimental results; and the last two sections discuss and conclude this study.

\section{Methods}

\subsection{The hierarchical feature fusion framework and its implementation}
\begin{figure*}[!h]
\centering
\includegraphics[width=4.0in,height=4in]{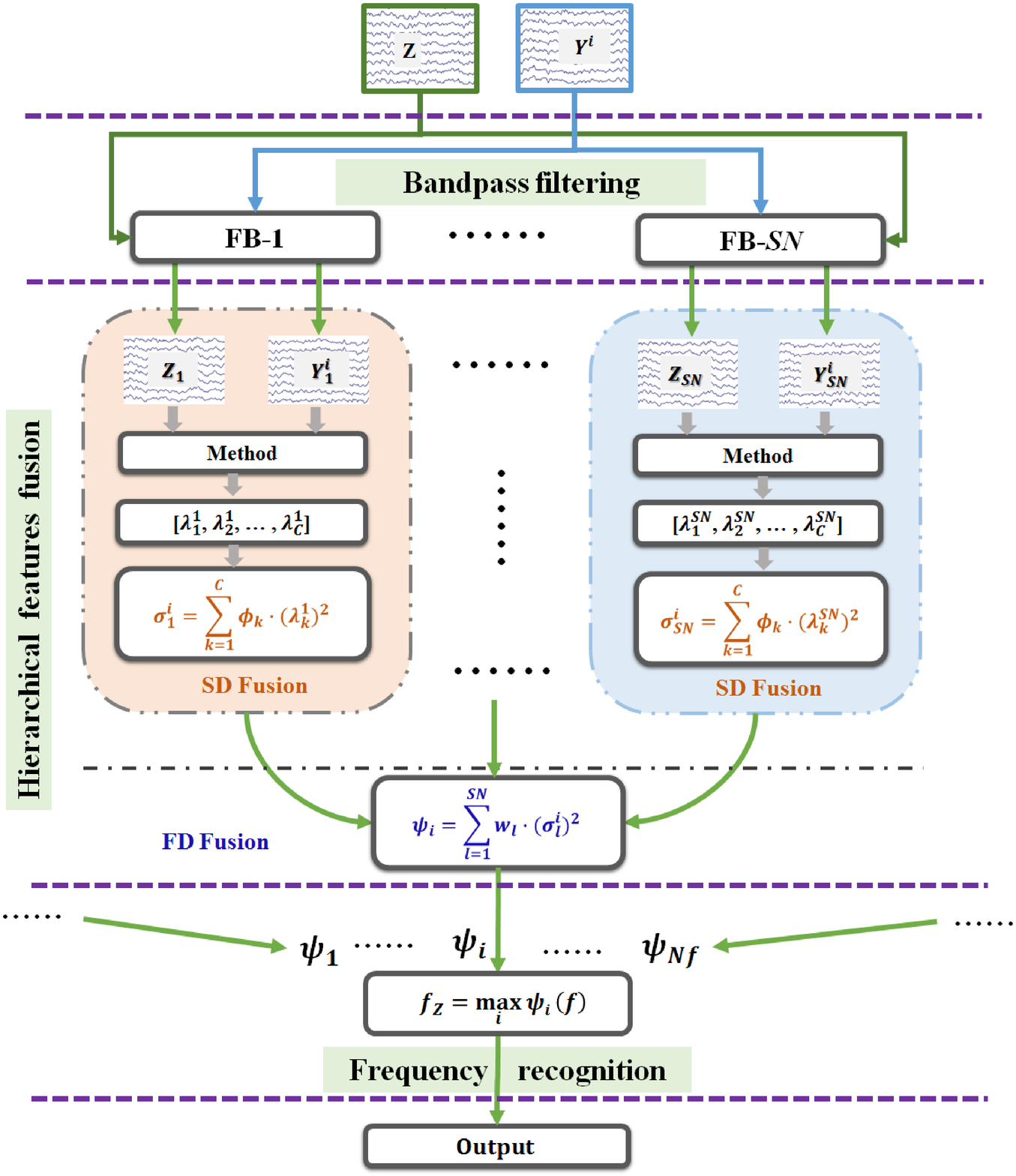}
\caption{Diagram of hierarchical feature fusion framework for frequency recognition. For a test sample $\bm{Z} \in \bm{R}^{C \times N}$ and individual template signals $\bm{Y}^{i}\in \bm{R}^{C \times N}$ at the $i$-th frequency $(i=1,2,\cdots,N_f)$, we first divide them to $SN$ sub-band signals using bandpass filters: $\bm{Z}_{1},\bm{Z}_{2} ,\cdots,\bm{Z}_{SN}$, $\bm{Y}^{i}_{1},\bm{Y}^{i}_{2},\cdots,\bm{Y}^{i}_{SN}$, respectively. In each sub-band, we can obtain $C$ features in descending order by a spatial filtering method, e.g., $ {\lambda}^{1}_{1}, \,\, {\lambda}^{1}_{2}, \,\, \cdots\,\, {\lambda}^{1}_{C}$ and $ {\lambda}^{SN}_{1}, \,\, {\lambda}^{SN}_{2}, \,\, \cdots\,\, {\lambda}^{SN}_{C}$ in the first and $SN$-th sub-bands. In SD fusion step, we fuse these features with the weights $\phi_{1},\phi_{2}, \cdots,\phi_{C}$  in each sub-band to produce new features: ${\delta}^{i}_{1}= \sum\limits_{k=0}^{C} \phi_{k} \cdot {\lambda}^{1}_{k}, \,\,{\delta}^{i}_{2}= \sum\limits_{k=0}^{C} \phi_{k} \cdot {\lambda}^{2}_{k}, \cdots,\,\,{\delta}^{i}_{SN}= \sum\limits_{k=0}^{C} \phi_{k} \cdot {\lambda}^{SN}_{k}$. Then, in the FD fusion step, the final feature at the $i$-th frequency for the test sample $\bm{Z}$ and the $\bm{Y}^{i} $ can be calculated as $\psi_{i}= \sum\limits_{l=0}^{SN} w_{l} \cdot {\delta}^{i}_{l}$. Finally, the frequency of $\bm{Z}$ is detected by the rule defined by formula (\ref{recogfrule1}). }
\label{fig:fig1}
\end{figure*}

The hierarchical feature fusion framework includes three main stages, i.e., bandpass filtering, hierarchical feature fusion and frequency recognition (see Fig. \ref{fig:fig1}).
\begin{itemize}
	\item In the stage of bandpass filtering, several bandpass filters are employed to filter the EEG signals, respectively, into multiple sub-band components. The filter bank design for the bandpass filters should be optimized according to the frequencies of the SSVEP stimulus. Here, all the sub-bands cover multiple harmonic frequency bands with the same high cut-off frequency at the upper bound frequency. for a more detailed analysis for the frequency band selection refer to ~\cite{chen2015filter}.
	\item  In the second stage, the hierarchical feature fusion includes SD fusion and FD fusion, respectively. First, for each pair of sub-band signals, a spatial filtering method yields multiple features that measure the correlation levels. Then, SD fusion is implemented on these features to obtain a new feature for each sub-band. Subsequently, FD fusion is used to generate the final features at each stimulus by combining the new features obtained by SD fusion on all the sub-bands. For both SD and FD fusions, many strategies could be used to fuse the features. In the present study, we used a weighted method with a nonlinear function to combine all the features obtained via the spatial filtering method. For the FD fusion method, we also implemented a weighted strategy with nonlinear function to combine all the features.
	\item  In the third stage, the proposed framework performs the frequency recognition based on the features obtained in the second stage. Supervised learning methods, such as linear discriminant analysis (LDA), deep learning, as well as unsupervised learning methods can be used in this stage. Here, we employ the latter because of its satisfactory performance in previous studies~\cite{Lin2007,chen2015PNAS,zhangys2018b}. Specifically, the target frequency can be recognized as that with the highest magnitude of final feature.
\end{itemize}

In the implementation of the framework, no prior knowledge about the weights can be directly inferred in the SD fusion step, and the weights themselves might depend on the chosen spatial filtering method. In this subsection, these weights are denoted as $\phi_{1},\phi_{2}, \cdots,\phi_{C}$.

Denote a test sample as $\bm{Z} \in \bm{R}^{C \times N}$ and the $i$-th frequency template signal as $\bm{Y}^{i}\in \bm{R}^{C \times N} (i=1,2,\cdots,N_f)$. Furthermore, denote the sub-band signals of $\bm{Z}$ and $\bm{Y}^{i}$  after the $l$-th bandpass filtering as $\bm{Z}_{l} \in \bm{R}^{C \times N}$ and $\bm{Y}^{i}_{l}\in \bm{R}^{C \times N}$ respectively, $l=1,2,\cdots,SN; \,\, i=1,2,\cdots,N_f$. With a spatial filtering method, $C$ features in descending order, denoted as $\left[\, {\lambda}^{l}_{1}, \,\, {\lambda}^{l}_{2}, \,\, \cdots\,\,,  {\lambda}^{l}_{C} \,\right]$, on the $\bm{Z}_{l}$ and $\bm{Y}^{i}_{l}$  can be obtained. Here, $C$ is the number of variables, $N$ is the number of samples, $N_f$ is the number of stimulus frequencies, and $SN$ is the number of bandpass filters.

Then with the SD fusion, the new feature in $l$-th sub-and at the $i$-th frequency can be obtained as :

\begin{equation}
{\delta}^{i}_{l}= \sum\limits_{k=1}^{C} \phi_{k} \cdot {\lambda}^{l}_{k}, \,\, l=1,2,\cdots,SN, \,\, i=1,2,\cdots,N_f. \label{fusingSD}
\end{equation}

With FD fusion, the final feature for the test sample $\bm{Z}$ and the template signal $\bm{Y}^{i}$ at the $i$-th frequency $(i=1,2,\cdots,N_f)$ can be obtained as :

\begin{equation}
\psi_{i}= \sum\limits_{l=1}^{SN} w_{l} \cdot {\delta}^{i}_{l}, \,\, i=1,2,\cdots,N_f. \label{fusingFD}
\end{equation}
where $w_{l},w_{2},\cdots,w_{SN}$ are the weights. They can be obtained by a weighting nonlinear function as below~(refer to ~\cite{chen2015filter}):

\begin{equation}
w_{m}= m^{-a_1}+b_1  \,\,,m=1,2,\cdots,SN.\,\, \label{weightF}
\end{equation}
where $a_1$ and $b_1$ are two parameters that control the classification performance, respectively.

Then, the frequency $f_{\bm{Z}}$ of  $\bm{Z}$ is specified as the frequency of the template signal that has maximal feature with $\bm{Z} $, as below:
 \begin{equation}
f_{\bm{Z} }=\max_f \psi_{i}(f)  ,  \; i= 1, 2,\ldots,N_{f} \,\, \label{recogfrule1}
\end{equation}

\subsection{The application of the framework on standard CORRCA method}
CORRCA method is developed based on the COCA, which is a technique to maximize the Pearson Product Moment Correlation coefficient between two multi-dimensional signals~\cite{Dmochowski2012}. Compared to CCA, COCA relaxes the constraint on orthogonality among the projection vectors, and generates a single projection vector for the two multi-dimensional input signals. Mathematically, COCA is an optimization problem, and its projection vectors can be obtained by solving a generalized eigenvalue problem. COCA has been used to investigate cross-subject synchrony of neural processing \cite{CohenENEURO2016}, and inter-subject correlation in evoked encephalographic responses \cite{dmochowski2014audience,Ki2016}. Recently, it was introduced for frequency recognition in SSVEP-based BCIs~\cite{zhangys2018b}. Below we provide a brief description of the standard CORRCA method.

Denote $\bm{X} \in \bm{R}^{C \times N} $ and  $\bm{Y} \in \bm{R}^{C \times N}$ as two multidimensional variables, where $C$ is the number of variables and $N$ the number of samples. COCA  seeks to find a projection vector $\bm{w}\in \mathbb{R}^{C \times 1}$ such that the resulting linear combinations  $\bm{x}=\bm{w}^{T}\bm{X}$ and $\bm{y}=\bm{w}^{T}\bm{Y}$  exhibit maximal correlation.

\begin{equation}
\begin{aligned}
\rho&=\arg\max_{\bm{w}}\frac{{\bm{x}}{\bm{y}^{T}}} {\left\| \bm{x} \right\| \ \left\| \bm{y} \right\|}\\
&=\arg\max_{\bm{w}} \frac{\bm{w}^{T} \bm{R}_{12} \bm{w}} {\sqrt{\bm{w}^{T}\bm{R}_{11}\bm{w}} \sqrt{\bm{w}^{T}\bm{R}_{22}\bm{w}}}\\
\end{aligned}
\,\, \label{corrca}
\end{equation}
where $\rho$ denotes the correlation coefficient. $\bm{R}_{11}=\frac{1}{N}{\bm{X}}{\bm{X}}^{T}$, $\bm{R}_{22}=\frac{1}{N}{\bm{Y}}{\bm{Y}}^{T}$, $\bm{R}_{12}=\frac{1}{N}{\bm{X}}{\bm{Y}}^{T}$, and $\bm{R}_{21}=\frac{1}{N}{\bm{Y}}{\bm{X}}^{T}$ are four sample covariance matrices. Supposing that $\bm{w}^{T}\bm{R}_{11}\bm{w} = \bm{w}^{T}\bm{R}_{22}\bm{w}$, differentiating formula (\ref{corrca}) with respect to $\bm{w}$ and setting to zero, we obtain the following eigenvalue equation \cite{Dmochowski2012}:

\begin{equation}
(\bm{R}_{12}+\bm{R}_{21}) \bm{w} =\lambda (\bm{R}_{11} + \bm{R}_{22}) \bm{w}
\,\, \label{corrca2}
\end{equation}

With formula (\ref{corrca2}), we can obtain $C$ correlation coefficients,  i.e., $\rho_{1},\rho_{2},\cdots,\rho_{C}$ for $\bm{X}$ and $\bm{Y}$. In the standard CORRCA method, only the maximal coefficient among $\{ \rho_{1},\rho_{2},\cdots,\rho_{C} \}$  is used as the  feature for  frequency recognition.

Suppose that $\bm{\bar{Z}} \in \bm{R}^{C \times N}$ is a test sample, $\bm{\bar{Y}}_{i}\in \bm{R}^{C \times N} $ is an individual template signal calculated by averaging SSVEP data across multiple trials at frequency $f_i$, $i=1,2,\cdots,N_f$. Denote maximal correlation coefficients between $\bm{\bar{Z}} $ and $\bm{\bar{Y}}_{i} $ as $\beta_i$, $i=1,2,\cdots,N_f$. Then, the frequency of  $\bm{\bar{Z}}$ is determined by finding the frequency of the template signal that has maximal correlation with $\bm{\bar{Z}}$, as below:
 \begin{equation}
f_{\bm{\bar{Z}}}=\max_f \beta_{i}(f)  ,  \; i= 1, 2,\ldots,N_{f} \,\, \label{recogfrule}
\end{equation}

\begin{figure*}[!h]
\centering
\includegraphics[width=3.5in,height=2.5in]{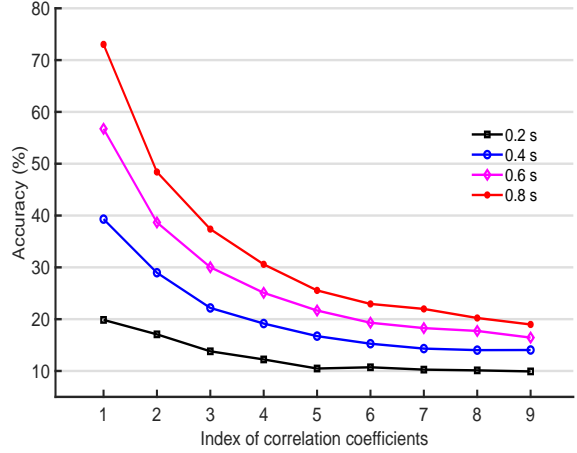}
\caption{Accuracy obtained by using each correlation coefficient of the CORRCA method. Four time windows (i.e., 0.2, 0.4, 0.6, and 0.8 s) were used for corresponding to the four series. The accuracies decrease nonlinearly as the correlation coefficient index increases.}
\label{fig:fig2}
\end{figure*}

Although the CORRCA method yields significantly better performance compared to the state-of-art CCA method, only the largest correlation coefficients in the set of $\{ \rho_{1},\rho_{2},\cdots,\rho_{C} \}$  is retained and the others are optionally discarded. In our preliminary analysis on the benchmark dataset, we investigated the classification performance of each correlation coefficient in the CORRCA method. As shown in Fig.\ref{fig:fig2},  the correlation coefficient index versus the averaged classification accuracy across subjects at four time windows is presented. Notice that almost all the coefficients could provide discriminative information for achieving classification accuracies above chance level ($1/40$). Therefore, exploring the strategy of combining all the correlation coefficients, namely through SD fusion, could be beneficial for enhancing the overall performance of the CORRCA method. In addition, the previous study demonstrated that the FD fusion can improve the performance of CORRCA method~\cite{zhangys2018b}. Accordingly, we believe that the proposed framework could further enhance system performance compared to using either SD fusion or FD fusion alone. In the current study, the CORRCA method was used as a representative method to implement the proposed framework, and investigate how its performance is affected.

In implementing the framework on the CORRCA method, no prior knowledge about the weights in the SD fusion step can be directly inferred.  As such, the key question optimal design of feature weights, i.e., the correlation coefficients. Intuitively, the discriminative ability should vary among the features, and the accuracy curve obtained with differing relative feature weights should be nonlinear. As shown in Fig.\ref{fig:fig2},  we find that the system accuracy decreases nonlinearly from the largest correlation coefficient to the smallest one, and generally the larger coefficients yield higher accuracies. The accuracy curve most closely follows an exponential function (as seen in Fig.\ref{fig:fig2}). Accordingly, we adopted a weighting nonlinear function to calculate the weights in the SD fusion, as below.

\begin{equation}
\phi_{k}=  {e}^{(-a_2 \cdot k)}+b_2  \,\,,k=1,2,\cdots,C.\,\, \label{weightlabel}
\end{equation}
where $a_2$ and $b_2$ are two parameters that control the classification performance. In current study, $a_2$ and $b_2$ were optimized via a grid search using a standard CORRCA method on the training set. Fig. 3 depicts the parameter selection process in which $a_2$ and $b_2$ were limited to [0:0.2:2] and [0:0.2:2], respectively. Here, the weights were calculated using formula (\ref{weightF}) with $a_0$ and $b_0$ specified as default values of 1.25 and 0.25, respectively. The classification accuracy and ITR with different $a_2$ and $b_2$ values with a time window of 0.8 s were calculated. The parameter values of $a_2$ and $b_2$ that led to the highest ITR were selected (i.e. 0.6 and 0), respectively. Then these values were applited onto the test set in the following calculation and analysis. More details of the corresponding procedure and parameter settings employed for bandpass filtering can be found in our previous study~\cite{zhangys2018b}. Note that the choice of formula (\ref{weightlabel}) is not unique, other function based may also be adequate. The CORRCA method with the proposed framework is termed as the hierarchical feature fusion CORRCA (HFCORRCA) hereafter.

\begin{figure*}[!h]
\centering
\includegraphics[width=4.5in,height=2.0in]{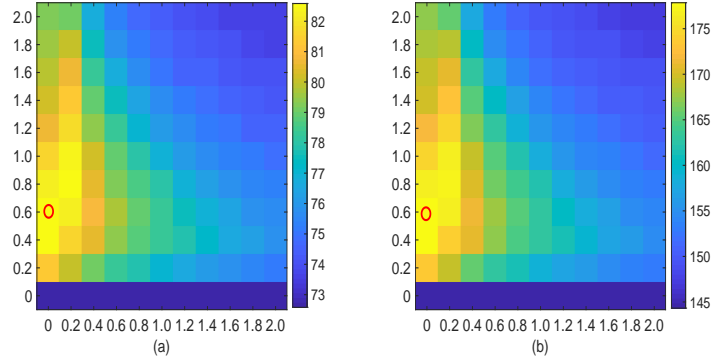}
\caption{Parameter selection process for $a_2$ and $b_2$. The averaged accuracy (\%) and ITR (bits/min) obtained by different combination of $a_2$ and $b_2$. The time window was set to 0.8 s. (a) accuracy, (b) ITR. The red circle indicates the optimal combination of $a_2$ and $b_2$ to yield the maximal accuracy. }
\label{fig:fig3}
\end{figure*}

\section{Experimental Study}
\subsection{EEG dataset}
The EEG dataset used for evaluation is a publicly available benchmark dataset consisting of offline SSVEP-based BCI spelling experiments on thirty-five healthy subjects (seventeen females, mean age 22 years)~\cite{wangyj2016}. In the BCI system, 40 frequencies (8-15.8 Hz with an interval of 0.2 Hz) were used for target coding. During the experiments, each participant was asked to spell for six blocks. In each block, 40 trials corresponding to all 40 stimulus frequencies were provided in a random order. Each trial lasted for 6 s, including 0.5 s for the visual cue and 0.5 s for the inter-stimulus interval.

EEG data were recorded at a sampling rate of 1000 Hz via a Synamps2 system (Neuroscan, Inc.). A total of 64 electrodes were placed according to the 64-channel extended international 10-20 standard. The EEG data were treated with a 50-Hz notch filter and subsequently with a 0.15-200-Hz band-pass filter. Then, each 6-s segment beginning at 0.5 s pre-stimulus was extracted. After that, the epoch was downsampled to 250 Hz. A more detailed description of the EEG dataset can be found in the literature \cite{wangyj2016}.

In current study, we divided the dataset into two parts, i.e. a training set and testing set. The data from first fifteen subjects were used as training set, and the remaining twenty subjects were used as the testing set.

\subsection{Performance evaluation}
In this study, we carry out an extensive comparison between the spatial filtering method with and without the proposed framework. For the spatial filtering method, CORRCA was used because of its robust performance. The classification accuracy and ITR were adopted as the evaluation metric. The 0.5 s cue time was considered in the calculation of ITR. In the current study, the individual templates of each frequency were generated by averaging SSVEP data across multiple blocks at the corresponding frequency. We adopted the leave-one-block-out cross validation method to evaluate the performance for the comparisons between the two methods. Specifically, one EEG block from a group of five blocks was selected as the training set and the remaining blocks were treated as the testing set. For each subject, this procedure was repeated six times such that the samples of each block were used as the testing set once.

Furthermore, we used the $r$-square value to evaluate the discriminability of the features obtained by each method. In the current study, the $r$-square value was computed using feature values of the attended target stimulus and the maximal feature values of the non-attended stimuli~\cite{nakanishi2015,zhangys2018b}, as the following formula~\cite{blankertz2007}.

\begin{equation}
R =\frac{\sqrt{N_1 \cdot N_2}}{N_1+N_2} \cdot \frac{mean(F_1)-mean(F_2)}{std(F_1\cup F_2)} \,\, \label{rsquare}
\end{equation}
where $F_1$ and $F_2$ are the sets of features of attended target stimulus and the maximal feature values of the non-attended stimuli, respectively. $N_1$ and $N_2$
are the numbers of features in these two sets. As the sign of this difference is important, we adopt the signed $r$-square value calculated as follows~\cite{blankertz2007}:

\begin{equation}
r^{2} ={\rm Sign}(R ) \cdot R^{2}  \,\, \label{signrsquare}
\end{equation}

\section{Results}
\begin{figure*}[!h]
\centering
\includegraphics[width=5in,height=2in]{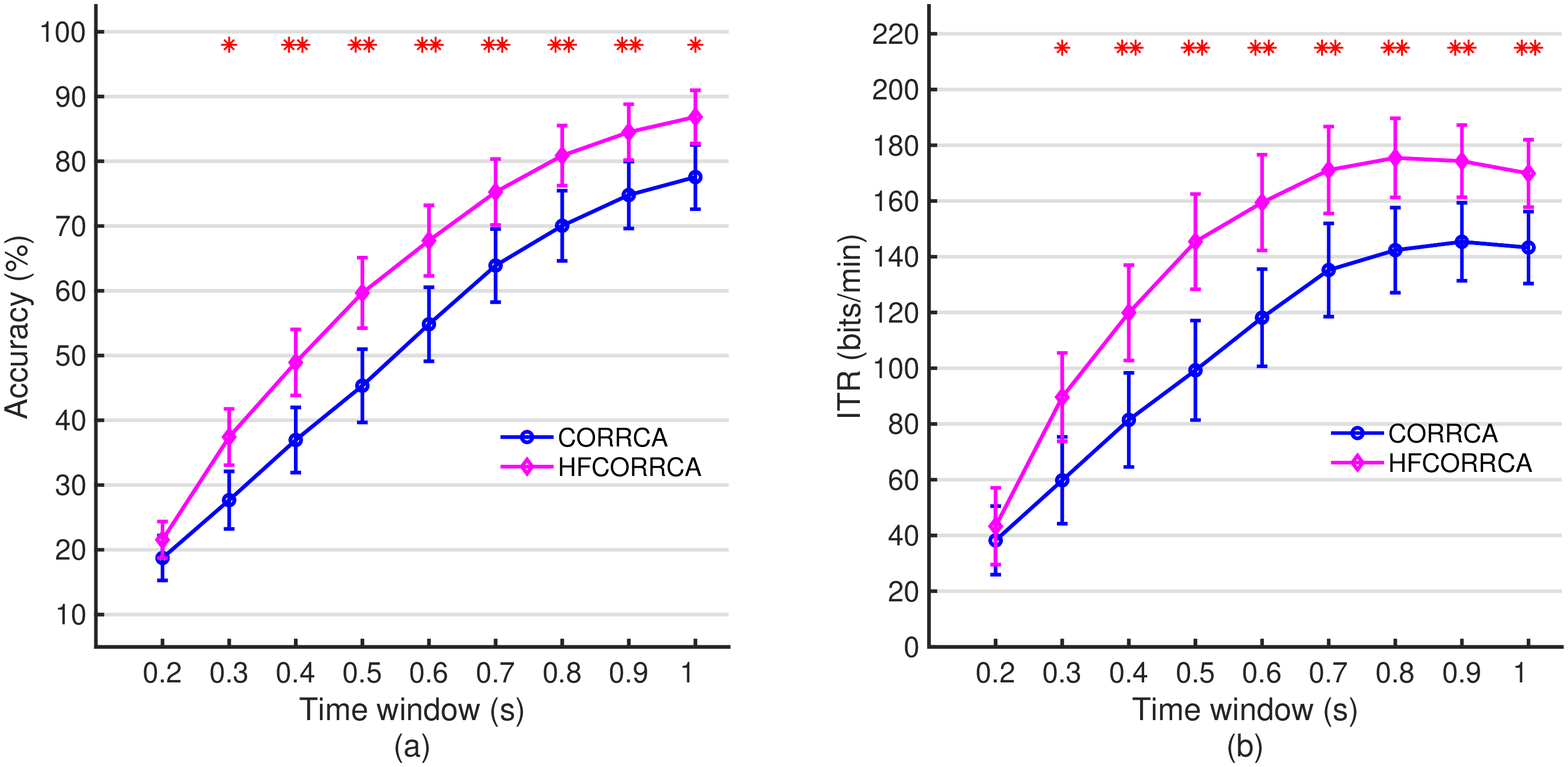}
\caption{Classification accuracies (a) and ITRs (b) averaged across subjects obtained by CORRCA and HFCORRCA at different time windows. Error bars indicate standard errors, *($p<0.05$) and ** ($p<0.001$) indicate significant difference between the two methods using paired $t$-tests.}
\label{fig:fig4}
\end{figure*}

Fig.\ref{fig:fig4} shows classification accuracies and ITRs averaged across all subjects at different time windows ranging from 0.2 s to 1 s with a step size of 0.1 s. As shown in Fig.\ref{fig:fig4}, the HFCORRCA method consistently outperforms the original CORRCA method in terms of accuracy and ITR, and their difference is significant when the time windows are above 0.2 s (paired $t$-tests). These results indicate that the proposed framework is feasible and can enhance the performance of the CORRCA method.

\begin{figure*}[!h]
\centering
\includegraphics[width=4.5in,height=2.6in]{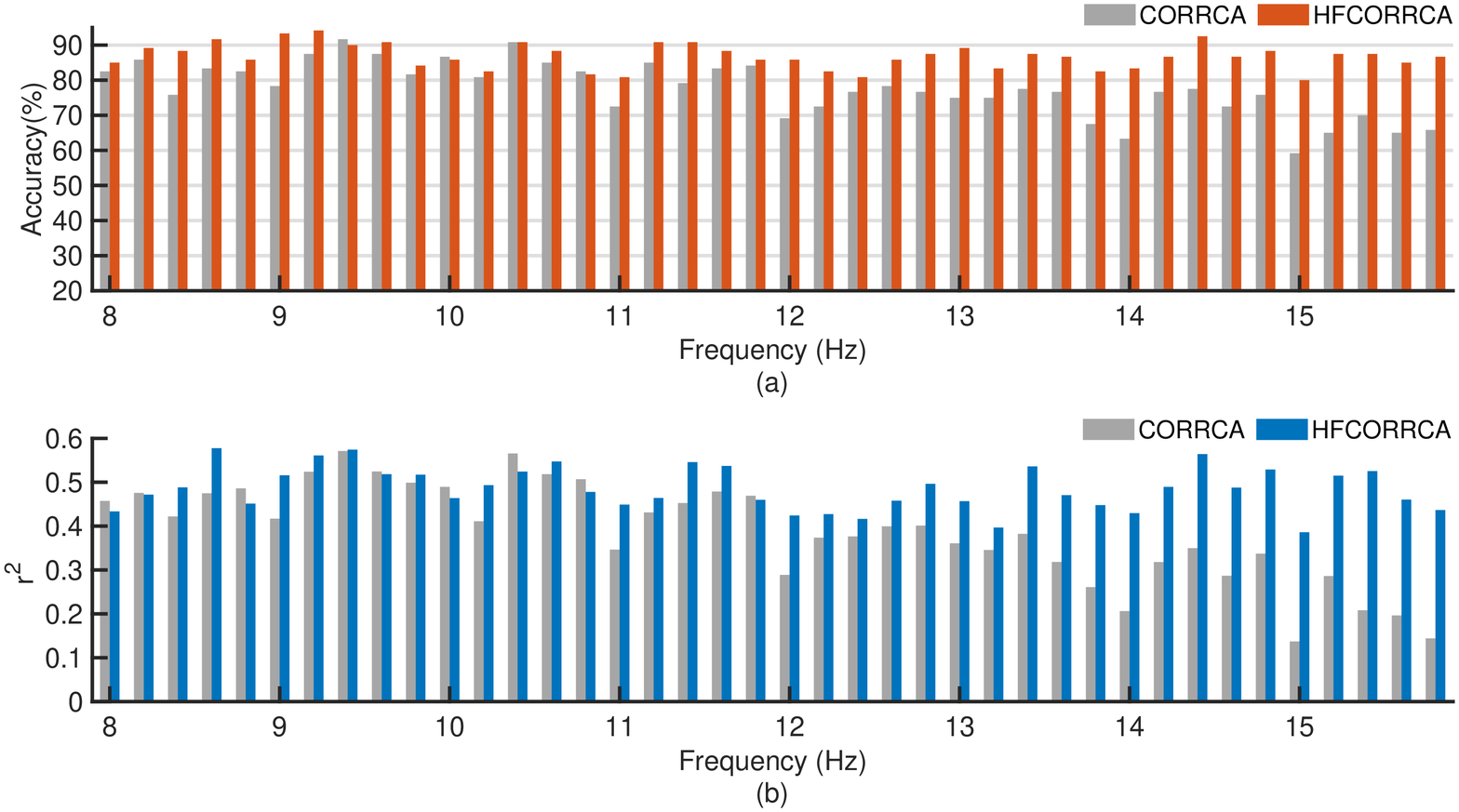}
\caption{Classification accuracies (a) and the $r$-square values (b) averaged across all subjects at the forty stimulus frequencies using a 1 s time window.}
\label{fig:fig5}
\end{figure*}

To further compare the performance between CORRCA and HFCORRCA methods, we present the averaged classification accuracies and the signed $r$-square values across all subjects at the 40 stimulus frequencies with a time window size of 1 s. As shown in Fig.\ref{fig:fig5}, the HFCORRCA method increased the vast majority of classification accuracies (37/40) and the $r$-square values (32/40) of the forty stimulus frequencies. Compared to the CORRCA, the standard deviation of classification accuracies using the HFCORRCA method decreased from 7.73 to 3.55, which indicates that HFCORRCA improves the reliability of frequency detection for all frequencies. In addition, the mean value of signed $r$-square values averaged across the forty frequencies increased from 0.39 to 0.48. These results further demonstrate that the proposed framework can produce more robust features compared to existing methods, and improve of the discriminability between attended target stimulus and non-attended non-target stimuli.

\begin{figure*}[!h]
\centering
\includegraphics[width=4.8in,height=2.1in]{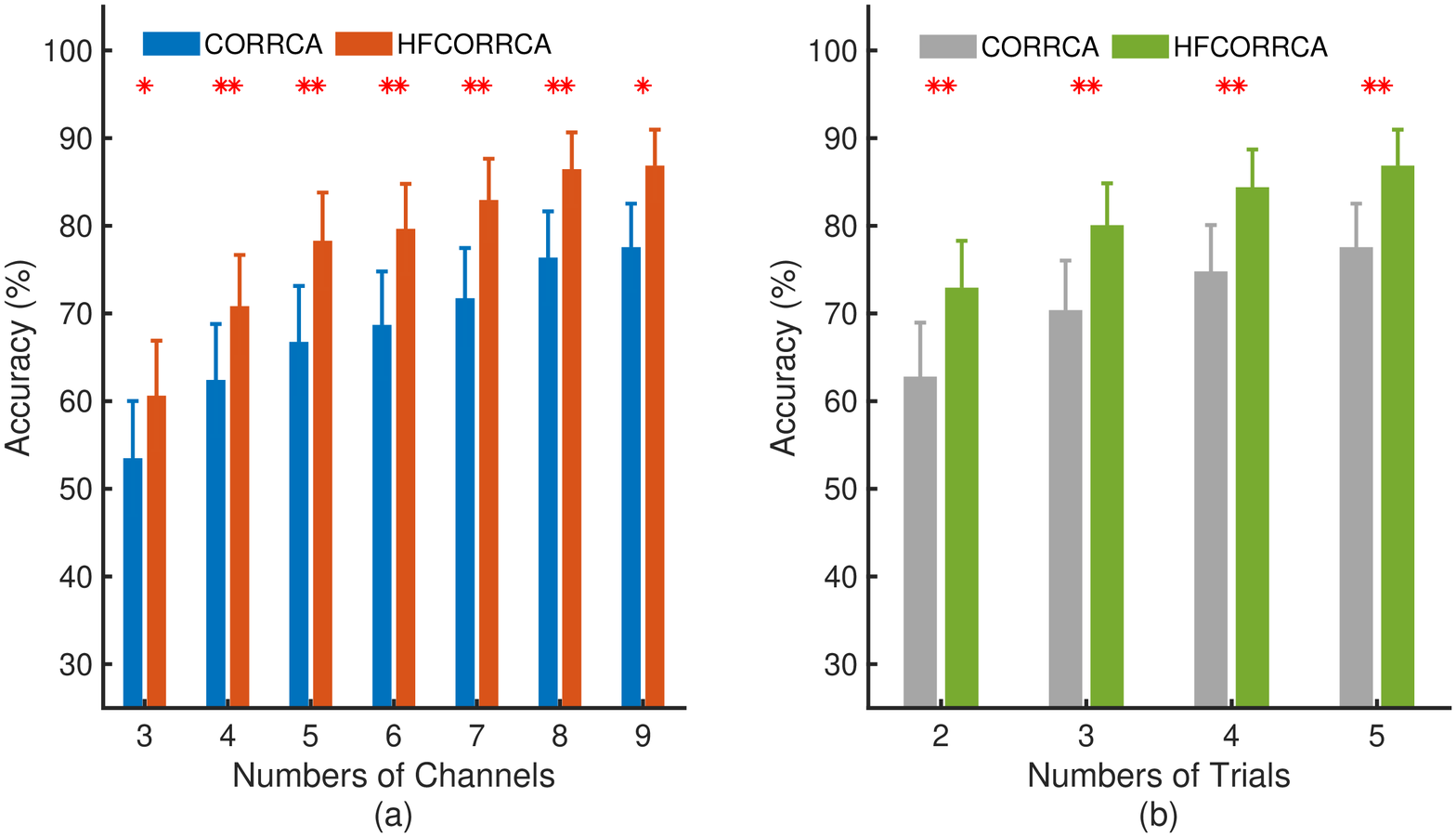}
\caption{ Classification accuracies averaged across all subjects obtained via CORRCA and HFCORRCA methods at different numbers of channels (a) and  different numbers of training trials,respectively with a time window of 1 s. Error bars indicate standard error, *($p<0.05$) and ** ($p<0.001$) of paired $t$-tests indicate significant difference between the two methods.}
\label{fig:fig6}
\end{figure*}

Previous studies have demonstrated that the number of channels and training trials can significantly influence performance~\cite{nakanishi2015,zhangys2018b}. Here we further investigate the classification accuracies of HFCORRCA and CORRCA methods at different number of channels and training trials. Fig.\ref{fig:fig6} illustrates the results of both methods using a 1-s time window. As shown, the HFCORRCA method yielded significantly better performance than the CORRCA method under all conditions. These results demonstrate that the proposed framework is a promising strategy to improve the performance of frequency recognition methods, such as the CORRCA method, to further enhance the performance of real-world systems.

\section{Discussion}
EEG is the most studied modality in BCI research, however, it is highly prone to be contamination by noise and artifacts. These factors usually distort the useful components in the signal and result in dispersion of the discriminative information across the projected signals or features~\cite{Lin2007,mishuhina2018}. Mishuhina {\em et al.} found that it is beneficial in motor imagery data processing to combine features obtained using CSP spatial filters~\cite{mishuhina2018}. Based on this prior research, we assume that fusing the features produced by different frequency recognition spatial filters can also provide more robust features and avoiding loss of the discriminative information for frequency recognition in SSVEP-based BCIs (SD fusion). For instance, the CORRCA method utilizes only the largest correlation coefficient corresponding to the spatial filter of the largest generalized eigenvalues, and all other features are discarded. Fusing all the correlation coefficients generated by all the spatial filters of CORRCA method can enhance overall performance. Additionally, filter bank methods are widely used in SSVEP-based BCI and motor imagery based BCIs to generate more discriminative features~\cite{chen2015filter, mishuhina2018,Zhangyu2017NS, zhangys2018b}. This technique achieves feature fusion in different frequency bands (FD fusion). Traditionally, these two types of features fusion methods are used independently in the BCI studies. To further boost the performance of SSVEP-based BCI systems, we propose a unified framework to integrate SD and FD fusion together for frequency recognition. The experimental results indicate that the framework can boost the frequency recognition when used with existing method, e.g., the CORRCA method, to enhance its performance. Specifically, the CORRCA method within the proposed framework significantly outperforms the contemporary methods under various verification conditions. Based on the proposed framework, some extended methods of spatial filtering are simply special cases of the proposed method. For instance,  filter bank CORRCA method is the HFCORRCA method without SD fusion.

\begin{figure*}[!h]
\centering
\includegraphics[width=3.3in,height=2.3in]{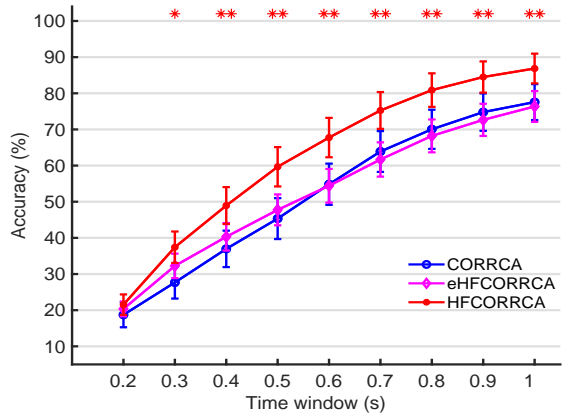}
\caption{Classification accuracies averaged across all subjects obtained by CORRCA, CORRCA with FD fusion and SD fusion using equal weights (eHFCORRCA), and HFCORCA, respectively, at different time windows. Error bars indicate standard error, *($p<0.05$) and ** ($p<0.001$) of paired $t$-tests indicate significant difference  between  eHFCORRCA and HFCORCA.}
\label{fig:fig7}
\end{figure*}

The standard CORRCA method can be considered within out framework by applying only the SD fusion on CORRCA method with a step function that has sharp transition from 1 to 0. This result of this is that only the largest coefficient is choosen. Conversely, for the HFCORRCA, we adopted a soft weighted function as formula (\ref{weightlabel}) to combine all the correlation coefficients. We kept all coefficients, and combined them with different weights. As shown Fig.\ref{fig:fig2}, we found that the accuracies decreased nonlinearly from the largest correlation coefficient to the smallest one. Intuitively, it is reasonable to use different weights on these coefficients. To verify this idea, as shown in Fig.\ref{fig:fig7}, we present the results obtained by the CORRCA, eHFCORRCA (a method that replaces the different weights used in HFCORRCA to equal weights, i.e., ${\phi}_1=1, {\phi}_2=1, \cdots, {\phi}_C=1$) and HFCORCA.  Although eHFCORRCA shows better performance than CORRCA at short time windows (less than 0.5 s), we find that using equal weights produces inferior results during fusion of features using the CORRCA method. These results further confirm the justification for adopting a nonlinear weighted  function to fuse the features. Note that the weighted function in formula (\ref{weightlabel}) can be changed to other functions, such as $w_k=k^{-a}+b (k=1,2,\cdots,C)$ in formula (\ref{weightF}). In future studies, we will investigate different weighting functions for different spatial filtering methods, which might further optimize system performance.

\begin{figure*}[!h]
\centering
\includegraphics[width=3.3in,height=2.5in]{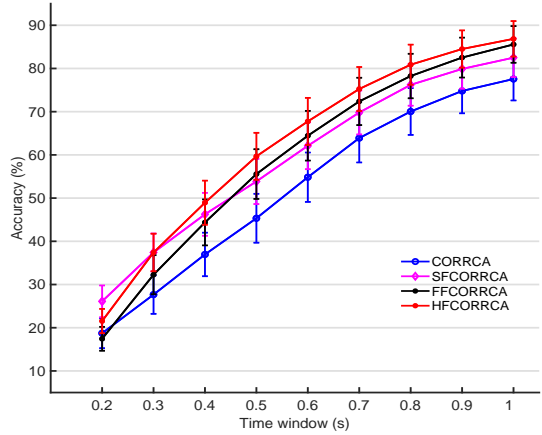}
\caption{The classification accuracies averaged across all subjects obtained by CORRCA, CORRCA with SD fusion(SFCORRCA),  CORRCA with FD fusion(FFCORRCA) and HFCORRCA, respectively, at different time windows. Error bars indicate standard error.}
\label{fig:fig8}
\end{figure*}

To confirm the contribution of each fusion operation (SD fusion and FD fusion) of the proposed framework, we calculated the classification accuracies using the CORRCA method with SD fusion only and FD fusion(FFCORRCA) only. As shown in Fig.\ref{fig:fig8}, although both SD fusion and FD fusion can improve the results compared to standard CORRCA, respectively, combining the two fusion methods resulted in the best performance (except at the shortest time window of 0.2 s). At the time window of 0.2 s, the CORRCA method with FD fusion produced lower accuracies than the original CORRCA method, which may be attributed to the short data length (50 samples) during bandpass filtering. It is believed that SD fusion in general is less sensitive to data length than FD fusion.

\begin{figure*}[!h]
\centering
\includegraphics[width=4.5in,height=2.3in]{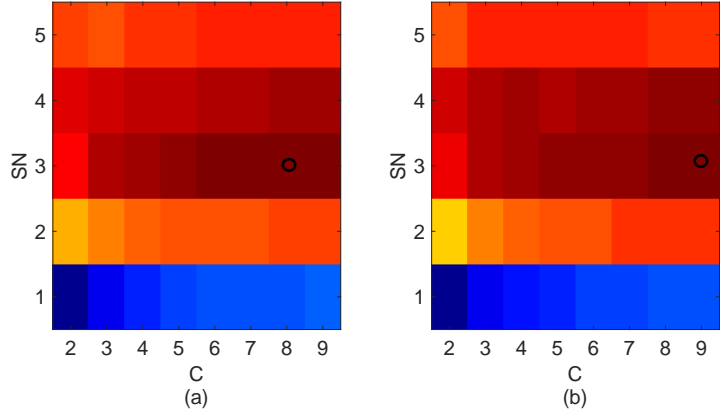}
\caption{Classification accuracies averaged across all subjects obtained by HFCORRCA with different combination of the number of features ($C$) and the number of subbands ($SN$) at two time windows.(a) 0.8 s and (b) 1 s. The black circle indicates the optimal values of $C$ and $SN$ to yield the maximal accuracy.}
\label{fig:fig9}
\end{figure*}

 In applying the fusion methods, we fused all the features in SD fusion step, and used five bandpass filters based on previous study \cite{chen2015filter}. We did not optimize the two parameters of $C$ and $SN$. As shown in Fig.\ref{fig:fig2}, the accuracies of the last four correlation coefficients are much lower than those of the first three correlation coefficients. One may argue that it is unnecessary to use all the correlation coefficients in the proposed framework. In order to provide some guidance for future online system development using HFCORRCA, we investigated how these two parameters influence the classification accuracies.  The $C$ and $SN$ were limited to [2:9] and [1:5], respectively, and two data time windows, i.e., 0.8 s, 1 s were used. The upper limit of $SN$ is equal to the number of channels in the EEG. As shown in Fig.\ref{fig:fig9}, for the two chosen time windows, the values of $SN$ which yielded the best results are same, but those of $C$ are different. We assessed the system accuracies when $C$ ranged from 4 to 9, and found that the accuracy differed by less than 0.5\% across the range. That is to say, the first four correlation coefficients can yield satisfactory results on the benchmark dataset. The optimal value of $C$ and $SN$ may depend on the specific dataset. For $SN$, when the stimulus frequencies belong to the same range used as those used in the benchmark dataset, it can be set to 3. For $C$, the simplest way is to set it equal to the number of the features produced by the SF method. When a calibration dataset is available, preliminary experiments can be used to optimize these values.

Lastly, it is worth mentioning that the proposed framework can only be applied directly on the standard frequency recognition methods (e.g., CCA, CORRCA) for now. In a future study, we will explore integrating this framework into extended frequency recognition methods, such as the two-stage method based on the standard CORRCA method (TSCORRCA)~\cite{zhangys2018b}. Moreover, the proposed framework can also be employed on CSP methods to boost classification rates in motor imagery based BCIs~\cite{mishuhina2018,Zhangyu2017NS}.

\section{Conclusion}
In summary, we propose a hierarchical features fusion framework to address the loss of discriminative information the occurs during feature extraction of spatial filtering methods. The framework consists of feature fusion using a nonlinear weighted function in both spatial and frequency domains. Under this framework, more robust features for frequency recognition are generated. Experimental results performed on a benchmark dataset demonstrated that the framework is effective at enhancing system performance. Specifically, and the improved CORRCA method within this framework significantly outperforms the original CORRCA method. This novel framework may be used to develop efficient methods for frequency recognition in SSVEP-based BCIs.

\section*{Acknowledgments}
This work was supported in part by the National Natural Science Foundation of China under Grant 61871423, Grant 61703407, Grant 81401484,  Grant 31771149, Grant 61522105, Grant 61527815, and Grant 61773129, in part by the Longshan academic talent research supporting program of SWUST under Grant 17LZX692. We also acknowledge Mr. Rami Saab for editing an early version of this manuscript.

\section*{References}


\end{document}